\documentclass{article}

\usepackage{spconf}
\usepackage{graphicx}
\usepackage{subfig, float}
\usepackage{color}
\usepackage{amsmath}
\usepackage{dsfont}
\usepackage{enumitem}
\usepackage{ragged2e}
\usepackage{comment}
\usepackage{hyperref}
\usepackage{multirow}
\usepackage{amssymb}
\usepackage{textcomp}

\newlength{\bibitemsep}\newlength{\bibparskip}
\let\oldthebibliography\thebibliography
\renewcommand\thebibliography[1]{%
  \oldthebibliography{#1}%
  \setlength{\parskip}{\bibitemsep}%
  \setlength{\itemsep}{\bibparskip}%
}
\begin{document}\sloppy

\title{Analyzing Real-Time Multimedia Content From Network Cameras\\Using CPUs 
and GPUs in the Cloud}

\name{Ahmed S. Kaseb$^1$, Bo Fu$^2$, Anup Mohan$^3$, Yung-Hsiang Lu$^2$, Amy Reibman$^2$, George K. Thiruvathukal$^4$}

\address{$^1$ Computer Engineering Department, Faculty of Engineering, Cairo University, Giza, Egypt \\ $^2$ School of Electrical and Computer Engineering, Purdue University, West Lafayette, IN, USA \\ $^3$ Intel Corporation, Santa Clara, CA, USA \\ $^4$ Department of Computer Science, Loyola University Chicago, Chicago, IL, USA \\ akaseb@eng.cu.edu.eg, fu200@purdue.edu, anup.mohan@intel.com, \\ yunglu@purdue.edu, reibman@purdue.edu, gkt@cs.luc.edu}

\maketitle

\begin{abstract}

Millions of network cameras are streaming real-time multimedia content (images 
or videos) for various environments (e.g., highways and malls) and can 
be used for a variety of applications.
Analyzing the content from many network cameras requires 
significant amounts of computing resources.
Cloud vendors offer resources in the form of cloud instances with different 
capabilities and hourly costs.
Some instances include GPUs that can accelerate analysis programs.
Doing so incurs additional monetary cost because instances with GPUs are more expensive.
It is a challenging problem to reduce the overall monetary cost of using the 
cloud to analyze the real-time multimedia content from network cameras while 
meeting the desired analysis frame rates.
This paper describes a cloud resource manager that solves this problem by 
estimating the resource requirements of executing analysis programs using CPU or 
GPU, formulating the resource allocation problem as a multiple-choice vector bin 
packing problem, and solving it using an existing algorithm.
The experiments show that the manager can reduce up to 61\% of the cost compared 
with other allocation strategies.
\end{abstract}


\section{Introduction}

\begin{figure}[!htbp]
\begin{tabular}{cc}

\subfloat[][University Lab]
{\includegraphics[width=1.55in]{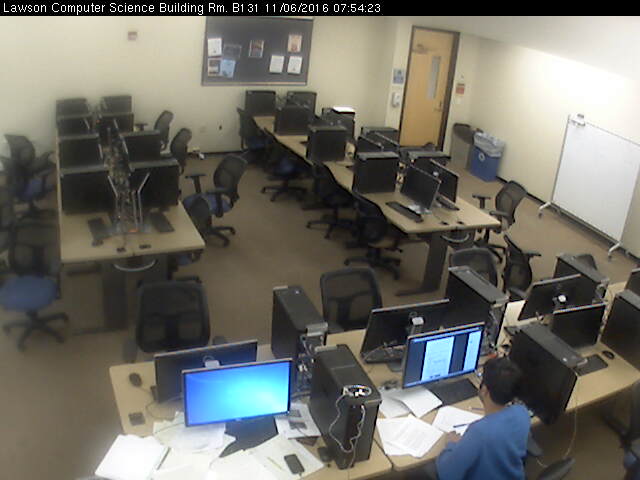}} &
\subfloat[][Park]
{\includegraphics[width=1.55in]{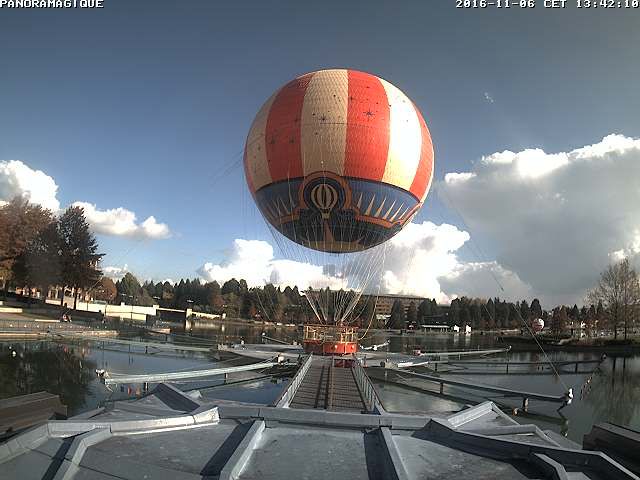}} \\
\subfloat[][Street Traffic]
{\includegraphics[width=1.55in]{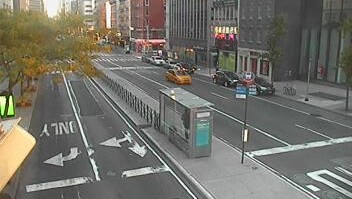}} &
\subfloat[][Houston Flood]
{\includegraphics[width=1.55in]{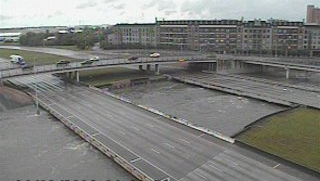}} 
\\
\end{tabular}
\caption{Images from network cameras.}
\label{fig:snapshots}
\vspace{-0.1in}
\end{figure}

Deployment of network cameras has been growing rapidly in recent years.
Network cameras stream real-time multimedia content (images or videos) that 
can be used for a variety of applications, for example, surveillance, 
entertainment, and traffic monitoring as shown in 
Figure~\ref{fig:snapshots}(a-c).
Analyzing the multimedia content from network cameras in real-time may also 
help first responders.
Figure~\ref{fig:snapshots}(d) shows an image from a camera during the 
Houston flood of April 2016.
Such multimedia content can be used to assess the severity of situations in 
different locations and to quickly respond to emergencies.

Analyzing the multimedia content from many network cameras requires a 
significant number of distributed computing resources.
Using the cloud can be beneficial because:
(i) Cloud vendors use a pay-as-you-go pricing model.
That means that users pay only when resources are used.
This may reduce the overall monetary cost if the analysis is needed 
occasionally (e.g., during emergencies).
(ii) Cloud vendors offer a variety of cloud instances with different 
capabilities and hourly costs.
Some instances include GPUs.
Using GPUs can accelerate analysis programs and achieve higher frame rates.
This incurs additional monetary cost because GPU instances are more expensive.
This variety makes it a challenging problem to meet the desired frame rates at 
the lowest possible cost.
This paper aims at solving this problem by introducing a resource manager 
that uses the GPU to achieve higher frame rates and considers both 
GPU and non-GPU instances to reduce the overall cost.
The manager conducts test runs to estimate the resource 
requirements of analysis programs.
The manager formulates the resource allocation problem as a multiple-choice 
vector bin packing problem to decide what instance 
types to use, how many instances to allocate, which data streams to assign to 
which instances, and which CPU or GPU to analyze the data streams.

To evaluate the manager, the experiments use two programs 
using convolutional neural networks (\mbox{VGG-16}~\cite{VGG16} and 
ZF~\cite{ZF}) to detect objects (e.g., persons) in images.
The experiments show that the manager can use the GPU to achieve speedup of
around 13 (or 16) for VGG-16 (or ZF) and also reduce the cost.
This paper has the following contributions:
\begin{itemize}[noitemsep,nolistsep]
\item It describes a resource manager that reduces the monetary cost of 
using cloud to analyze real-time multimedia content from network 
cameras while meeting the desired analysis frame rates.
\item The manager uses GPU to achieve higher frame rates and considers both GPU 
and non-GPU instances to reduce the overall cost.
\item The manager considers several factors while allocating resources: 
(i) the resource requirements of executing analysis programs on either the CPU 
or the GPU, 
(ii) the desired frame rates,
(iii) the sizes of the frames provided by the cameras, and
(iv) the types and costs of both the GPU and non-GPU instances.
\item The manager formulates the resource allocation problem as a 
multiple-choice vector bin packing problem and solves it using an 
existing algorithm.
The experiments show that the manager is able to reduce 61\% of the cost 
compared with other allocation strategies.
\end{itemize}
\section{Related Work}

The visual data from many network cameras is publicly available through 
many sources, such as Departments of Transportation (e.g., 
\url{http://www.ohgo.com/}).
This data can be used for many applications, such as weather detection
\cite{ChenYangLindnerBarrenetxeaVetterli2012HowisWeatherImageProcessing}
and surveillance~\cite{srivastava2013video}.
Several systems have been developed for analyzing the visual data from cameras, 
such as IBM Smart Surveillance System~\cite{Tian2008} and 
CAM$^2$~\cite{CAM2GlobalSIP}.

Zhu et al.~\cite{zhu2011multimedia} explained the advantages of using cloud 
for multimedia applications.
Kaseb et al.~\cite{CAM2CCBD} proposed a resource manager to reduce the cost of 
analyzing the data from cameras, but do not consider GPU 
resources.
GPUs can be used to accelerate general purpose computation, 
such as image processing and computer vision~\cite{4607358}.
Different studies used GPUs for face detection~\cite{7177522}, motion 
estimation~\cite{4284972}, body tracking~\cite{6011847}, etc.
This paper considers using GPUs to accelerate and reduce the monetary cost of 
analyzing the real-time multimedia content from network cameras using the cloud.

\section{The Cloud Resource Manager}

\begin{figure}[!t]
\centering
\includegraphics[width=0.98\linewidth]{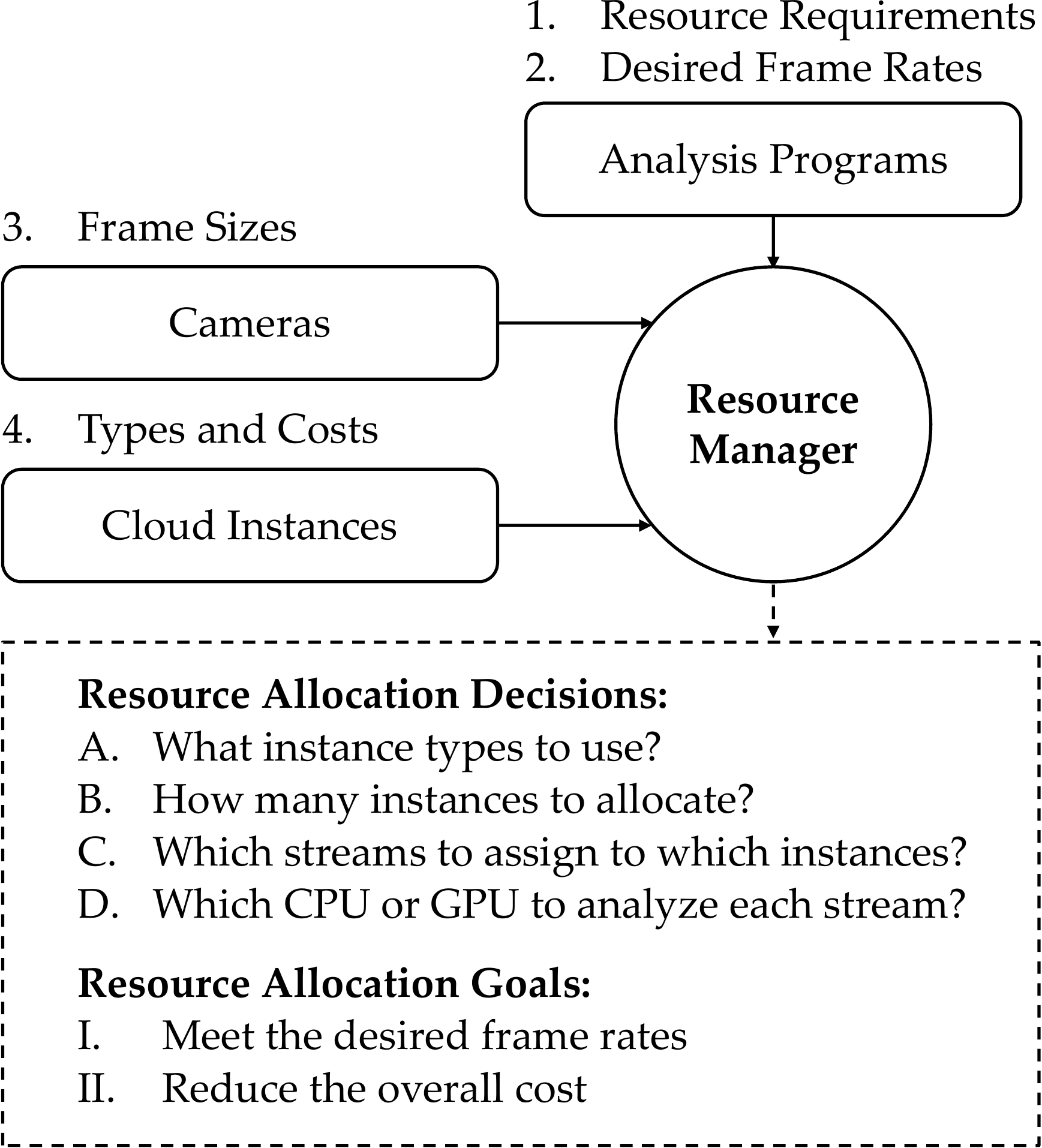}
\caption{The main factors (1-4) affecting resource allocation decisions (A-D) 
and the resource allocation goals (I and II).}
\label{fig:problem}
\vspace{-0.1in}
\end{figure}

The resource manager aims at meeting the performance requirements (i.e., meeting the desired frame rates) at the lowest possible monetary cost.
The performance of analyzing a single data stream is defined as the 
ratio between the actual analysis frame rate and the desired frame rate.
The overall performance of the system is then defined as the average 
performance for all the data streams.
The manager aims at maintaining the overall performance above 90\%.
Our experiments show that this can be achieved by maintaining the utilization 
of all the resources below 90\%.
The performance decreases if the resources are overutilized.
There is a clear trade-off between meeting the performance requirements and 
reducing the cost.
Allocating fewer instances than necessary decreases the performance, while 
allocating more increases the cost.
Figure~\ref{fig:problem} shows the main factors affecting resource allocation decisions as well as the resource allocation goals.
Section 3.1 discusses the factors considered by the resource manager.
Section 3.2 explains how the manager makes resource allocation decisions to achieve its goals.

\vspace{-0.1in}
\subsection{Factors Affecting Resource Allocation Decisions}

The resource manager considers the following factors while making 
allocation decisions:

\textbf{1. Resource Requirements:} The manager considers the following types of 
resources: CPU, memory, GPU, and GPU memory.
Different analysis programs require different amounts of resources.
For example, some programs are memory intensive while others are CPU intensive.
Moreover, some programs have implementations using GPU to achieve higher frame 
rates.
The resource requirements of these programs change according to which 
implementation is used (i.e., CPU or GPU).
The resource manager is designed to be used for a variety of applications.
Hence, it does not assume any prior knowledge about the analysis programs'
resource requirements.
The manager conducts two test runs (one using the CPU and the other using the GPU) to estimate the resource requirements of each program by monitoring the utilization of resources while executing the program.
The test runs are conducted once and the estimations of the resource 
requirements can be used for future executions of the same program.

\textbf{2. Desired Frame Rates:}
The frame rate at which an analysis program is executed significantly affects 
its resource requirements.
Experiments show that the CPU and GPU requirements of an 
analysis program increase linearly with its frame rate.
Using this linear relationship, the manager can estimate the resource 
requirements of an analysis program at different frame rates using a single test 
run conducted at a particular frame rate.
In addition, the frame rate may affect different types of resources differently.
For example, increasing the frame rate may increase its CPU 
requirement, but may have no effect on its memory requirement.
This causes some analysis programs to be CPU intensive at high frame rates 
while being memory intensive at low frame rates.

\textbf{3. Frame Sizes:}
Different cameras provide streams with different frame sizes (e.g., 
640$\times$480).
Higher frame sizes require higher resource requirements.
The effect of the frame size on the resource requirements of an analysis 
program depends on the time complexity and the space complexity of the program.
Since the resource manager assumes no prior knowledge about analysis 
programs, the manager repeats the test runs for each unique frame size.
Fortunately, there are only a few common frame sizes among network cameras.

\textbf{4. Types and Costs of Cloud Instances:}
Cloud vendors offer many instances with different capabilities and hourly costs.
Table~\ref{tab:gpu_instances} shows the capabilities and hourly costs of some Amazon EC2 instance types with and without GPUs.
The table shows that GPU instances (i.e., g2.2xlarge and g2.8xlarge) are 
more expensive than non-GPU instances (i.e., c2.2xlarge and c2.8xlarge).
The manager decides the types and number of instances needed to meet the 
desired frame rates at the lowest possible cost.

\begin{table}[!t]
\begin{center}
\caption{The capabilities and the hourly costs of some Amazon EC2 instance 
types with and without GPUs (at Oregon).}
\begin{tabular}{ | l | r | r | r | r |} \hline

\multicolumn{1}{|c|}{Instance} &
\multicolumn{1}{c|}{Cores} &
\multicolumn{1}{c|}{Memory (GB)} &
\multicolumn{1}{c|}{GPUs} &
\multicolumn{1}{c|}{Cost} \\ \hline
c4.2xlarge & 8 & 15 & - &\$0.419 \\ \hline
c4.8xlarge & 36 & 60 & - & \$1.675 \\ \hline
g2.2xlarge & 8 & 15 & 1 & \$0.650 \\ \hline
g2.8xlarge & 32 & 60 & 4 & \$2.600 \\ \hline
\end{tabular}
\label{tab:gpu_instances}
\end{center}
\vspace{-0.3in}
\end{table}

\subsection{Multiple-Choice Vector Bin Packing}

\begin{figure}[!t]
\centering
\includegraphics[width=0.98\linewidth]{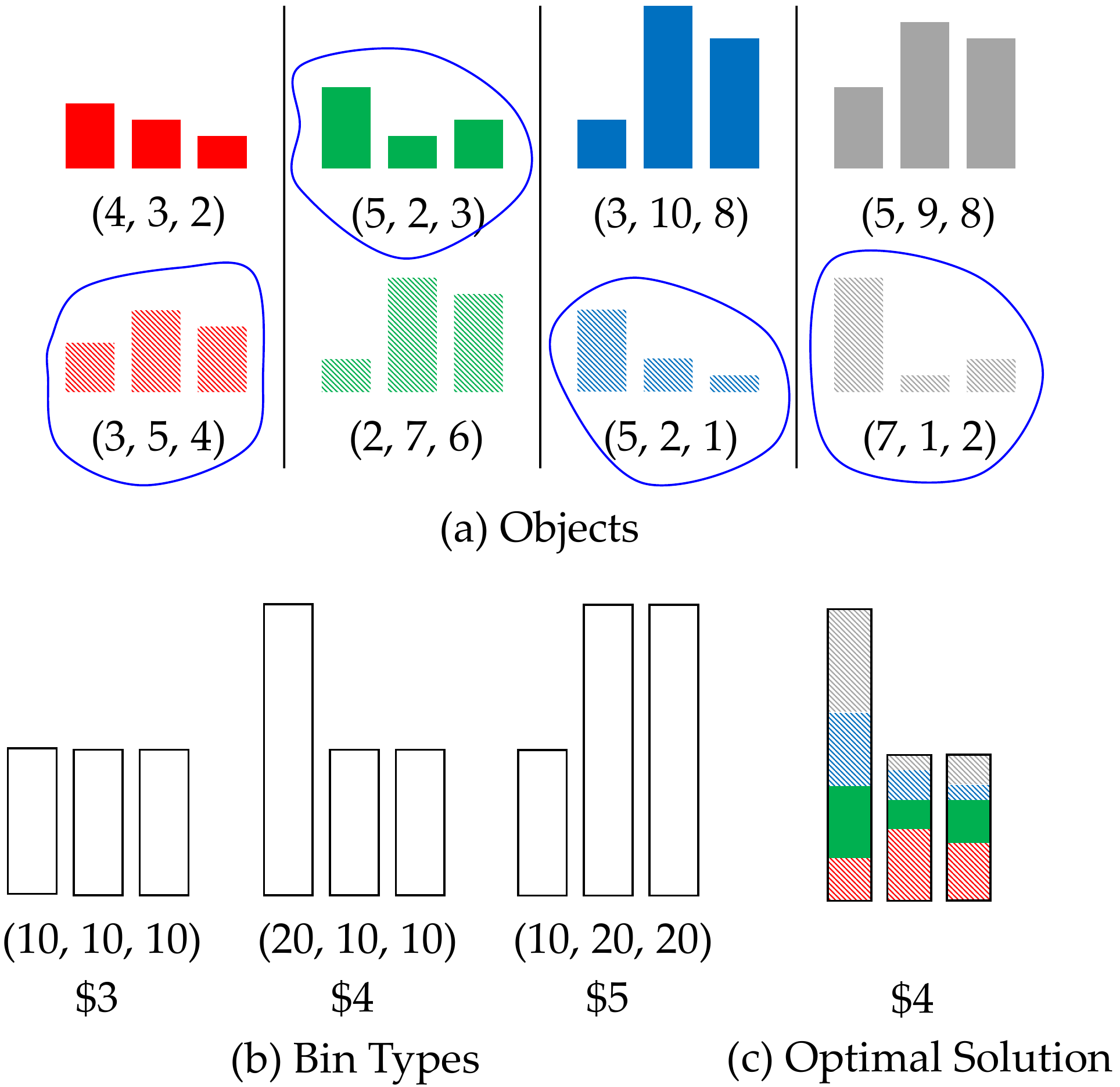}
\caption{An example of a multiple-choice 3D bin packing problem with: (a) four objects, each with two possible sizes (b) three bin types (c) the optimal solution using one bin containing all the four objects. This paper uses the algorithm proposed by Brandao and Pedroso~\cite{Brando201656} to solve the multiple-choice vector bin packing problem.
}
\label{fig:bin_packing}
\vspace{-0.1in}
\end{figure}

To make the resource allocation decisions shown in Figure~\ref{fig:problem}, this paper formulates resource allocation as a multiple-choice vector bin packing problem.
In this problem~\cite{Brando201656}, a bin has a cost and a multidimensional 
size.
An object may have one of several possible sizes (multiple choices).
The goal is to pack all the objects into bins such that:
(i) One size is selected for each object.
(ii) The overall cost of the used bins is minimized.
(iii) The total size of the objects in each bin does not exceed its size in 
any dimension.
Figure~\ref{fig:bin_packing} shows an example of a multiple-choice 3D bin packing problem.

Similarly, in the resource allocation problem, each instance has an hourly cost 
and a vector representing its resource capabilities (i.e., CPU, 
memory, GPU, and GPU memory).
For example, the vector $[8, 15, 0, 0]$ represents a non-GPU instance with 8~CPU 
cores, 15 GB of memory, and no GPUs (e.g., c4.2xlarge).
The vector $[8, 15, 1536, 4]$ represents a GPU-instance instance with 8~CPU 
cores, 15 GB of memory, and a single GPU with 1536 cores and 4 GB of memory (e.g., g2.2xlarge).
Each data stream may have one of two possible resource requirements depending 
on whether it is executed by the CPU or the GPU.
For example, the resource requirements of a program may be represented by the 
vector $[4, 0.75, 0, 0]$ or $[0.8, 0.45, 153.6, 0.28]$ if the program is 
executed by the CPU or the GPU respectively.
This means that if c4.2xlarge executes this program, the CPU utilization would 
be 50\% (i.e., 4/8).
If g2.2xlarge executes this program using the GPU, the CPU utilization would 
drop to 10\% (i.e., 0.8/8) and the GPU utilization would be 10\% 
(i.e., 153.6/1536).
The goal is to assign all the streams to instances such that:
(i) One resource requirement is selected for each stream.
This implies deciding if the stream is analyzed by the CPU or the GPU.
(ii) The overall cost of all the instances is minimized.
(iii) The total resource requirements of all the streams in each instance do 
not exceed the instance's capabilities for any resource type.

If instances with multiple GPUs (e.g., g2.8xlarge) are available, the dimensions 
and the multiple-choices of the problem change accordingly.
For example, the vector $[8, 15, 1536, 4, 1536, 4,$ $1536, 4, 1536, 4]$ 
represents an instance with 8~CPU cores, 15 GB of memory, and 4 GPUs each with 
1536 cores and 4 GB of memory (e.g., g2.8xlarge).
In this case, the vector $[8, 15, 0, 0, 0, 0, 0, 0, 0, 0]$ represents an 
instance with 8~CPU cores, 15 GB of memory, and no GPUs (e.g., c4.2xlarge).
Each data stream may have one of 5 possible resource requirements 
depending on whether it is executed by the CPU or one of the 4 GPUs.
In general, the dimension of the problem is $2 + 2 \times N$ where $N$ 
is the maximum number of GPUs in an instance.
That is because there are 2 resource types (i.e., CPU and memory) for the 
instance and 2 more resource types (i.e., GPU and GPU memory) for each added GPU.
The number of choices for the resource requirements of each stream is $1 + N$ 
because the stream can be analyzed either by the CPU or by one of the $N$ GPUs.

To solve the multiple-choice vector bin packing, the manager uses the
exact method proposed by Brandao and Pedroso~\cite{Brando201656} and provided 
through VPSolver (Vector Packing Solver, \url{http://vpsolver.dcc.fc.up.pt/}).
The output of the solver is the types and numbers of bins, which 
objects are assigned to each bin, and the selected size of 
each object.
In the resource allocation problem, this maps to the types and numbers of instances, which streams are assigned to each instance, and the 
selected resource requirement of each stream (i.e., which CPU or GPU to analyze 
the stream).
This output precisely represents the resource allocation decisions.
\section{Experiments and Results}

\subsection{Experimental Setup}
\label{sec:setup}

\begin{figure}[!t]
\begin{tabular}{cc}

\subfloat[][VGG-16]
{\includegraphics[width=1.55in]
{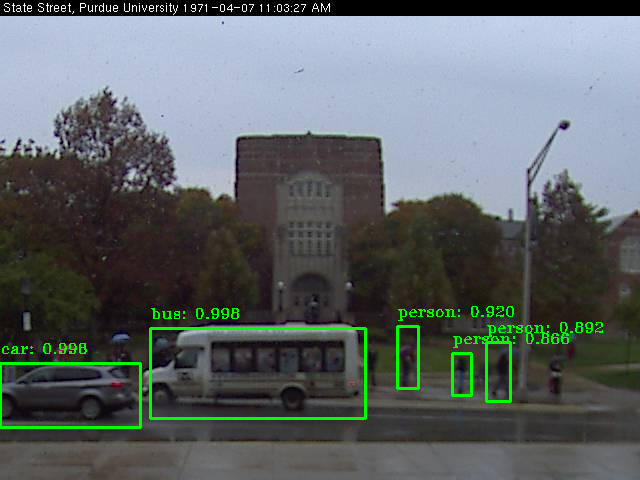}} &
\subfloat[][ZF]
{\includegraphics[width=1.55in]
{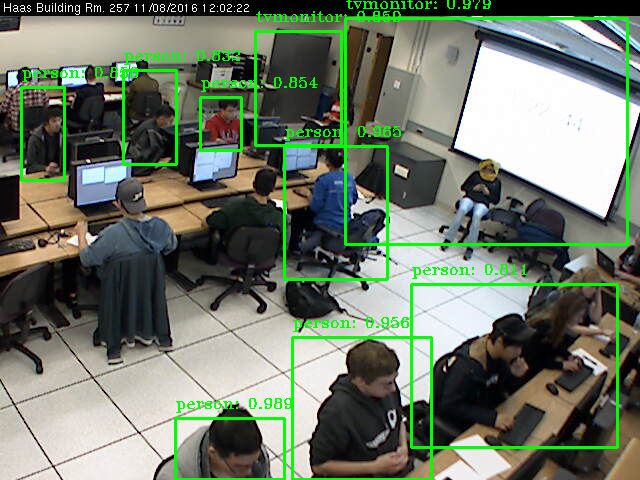}} \\
\end{tabular}
\caption{Sample output results from two network cameras. The objects 
detected are persons, cars, buses, and monitors.}
\label{fig:output}
\vspace{-0.1in}
\end{figure}

The experiments use two programs using convolutional neural networks 
(\mbox{VGG-16}~\cite{VGG16} and ZF~\cite{ZF}) to detect objects (e.g., persons) in images.
The experiments use the Python implementation of the region proposal network 
proposed by Ren et al.~\cite{FasterRCNN} to reduce the execution time of VGG-16 
and ZF.
Figure~\ref{fig:output} shows sample outputs.
All the experiments analyze 640$\times$480 MJPEG streams 
from network cameras.

The experiments use a machine with an 8-core Intel Xeon E5-2623 v3 CPU and 32GB 
of memory.
The machine has an NVIDIA K40 GPU with a 12GB of memory.
The experiments refer to the machine as a \emph{non-GPU instance} (or a 
\emph{GPU instance}) when the GPU is not used (or used) respectively.
The same pricing of the c4.2xlarge and g2.2xlarge instances (Table~\ref{tab:gpu_instances}) is used. 
The resource manager is generic and can be used with different cloud 
vendors with the appropriate changes in instance capabilities and hourly costs.
The experiments focus on the CPU and GPU utilization because the analysis programs are compute intensive, but the resource manager is generic and considers other resource types (e.g., memory and GPU memory).

\subsection{Speedup Using GPU}
\label{sec:speedup}

The main goal of the resource manager is to meet the desired frame rates of the analysis programs.
Using GPU allows the manager to accelerate the programs to achieve higher frame rates.
Table~\ref{tab:exp1} shows the effect of using the GPU on the maximum 
achievable frame rates of different analysis programs.
This shows that the manager can use the GPU to achieve a 
speedup of around 13 (or 16) for VGG-16 (or ZF).

\begin{table}[!t]
\renewcommand{\arraystretch}{1.3}
\caption{
The effect of using the GPU on the maximum achievable frame rates.
}
\label{tab:exp1}
\centering
\begin{tabular}{ | r | r | r | r |} 
\hline

\multirow{2}{*}{Program} & \multicolumn{2}{|c|}{Frame Rate (FPS)} & 
\multirow{2}{*}{Speedup}
\\ \cline{2-3}
& Using CPU & Using GPU & \\ \hline \hline

 VGG-16 & 0.28 & 3.61 & 12.89 \\ \hline
 ZF & 0.56 & 9.15 & 16.34\\ \hline

\end{tabular}
\vspace{-0.1in}
\end{table}

\subsection{Factors Affecting Resource Allocation Decisions}

\begin{figure}[!t]
\centering
\includegraphics[width=0.98\linewidth]{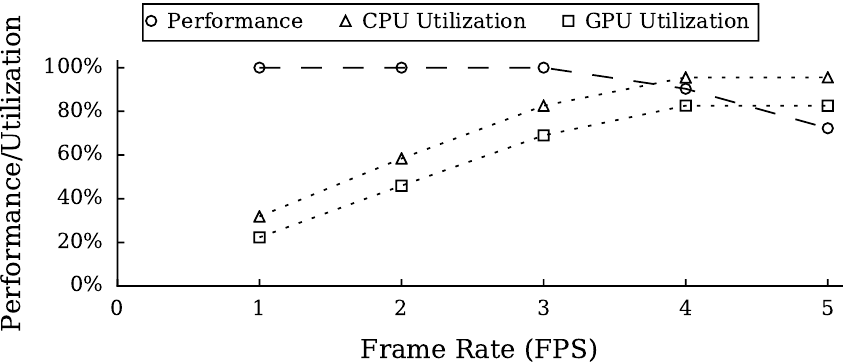}
\caption{
The effect of the desired frame rate on the resource requirements of 
VGG-16 as well as the analysis performance.
}
\label{fig:exp3_VGG-16}
\vspace{-0.1in}
\end{figure}

Desired frame rates significantly affect the resource requirements of analysis 
programs as well as the performance.
Figure~\ref{fig:exp3_VGG-16} shows this effect by executing VGG-16 using the 
GPU at different frame rates.
The figure shows that the CPU and GPU utilization increase 
linearly with the frame rate and the performance is 100\%.
The performance starts to drop gradually after the CPU resources get 
overutilized.
Since the resource manager aims at maintaining the analysis performance above 
90\%, the manager attempts to allocate cloud instances such that no resource 
utilization is above 90\%.

Table~\ref{tab:exp2} shows the CPU and GPU requirements of VGG-16 and ZF if 
executed at 0.2 FPS using the CPU or the GPU.
For each program, there are two choices 
of resource requirements depending on whether it is executed by the CPU or the 
GPU. The manager estimates these resource requirements for different frame 
rates based on a test run conducted at a particular frame rate  and the linear relationship between the frame rate and the CPU and GPU utilization shown in Figure~\ref{fig:exp3_VGG-16}.

\begin{table}[!t]
\renewcommand{\arraystretch}{1.3}
\caption{
The CPU and GPU requirements of VGG-16 and ZF if executed at 0.2 FPS using 
the CPU or the GPU.
}
\label{tab:exp2}
\centering
\begin{tabular}{ | r | r | r | r | r |} 
\hline

\multirow{2}{*}{Program} &
\multicolumn{2}{|c|}{Using CPU} &
\multicolumn{2}{|c|}{Using GPU} \\ \cline{2-5}

 & CPU & GPU & CPU & GPU \\ \hline \hline
 VGG-16 & 39.4\% & - & 5.3\% & 4.6\%\\ \hline
 ZF & 17.8\% & - & 2.2\% & 1.2\%\\ \hline

\end{tabular}
\vspace{0.1in}
\end{table}

\begin{figure}[!t]
\centering
\includegraphics[width=0.98\linewidth]{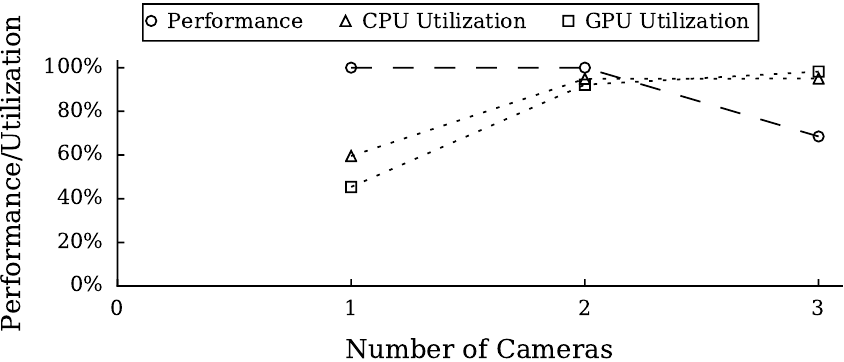}
\setlength{\belowcaptionskip}{-0.2in}
\caption{
The effect of the number of data streams being analyzed (using VGG-16 at 2 
FPS) on the resource utilization as well as the analysis performance.
}
\label{fig:exp4_VGG-16}
\end{figure}

The number of streams being analyzed using an instance 
affects its resource utilization and the performance.
Figure~\ref{fig:exp4_VGG-16} shows this by using the GPU to execute 
VGG-16 at 2 FPS on the streams from multiple cameras.
The figure shows that the CPU and GPU utilization increase 
almost linearly with the number of cameras and the performance is 100\%.
The performance drops after the CPU and GPU get overutilized.

\vspace{-0.1in}
\subsection{Evaluation of Resource Allocation}
\label{sec:resource_allocation}

To evaluate the allocation strategy of the manager described in this 
paper, we compare it with two different strategies as shown in Table~\ref{tab:strategies}.
All the strategies benefit from the ability of the manager to estimate the 
resource requirements of different analysis programs, to formulate the problem 
as a multiple-choice vector bin packing problem, and to solve it.
For ST1 (or ST2), there is a single choice for the resource requirements of 
each program because only non-GPU (or GPU) instances are considered.
The manager described in this paper uses ST3 which considers both non-GPU and 
GPU instances, hence, two choices of resource requirements exist for each 
program.

\begin{table}[!t]
\renewcommand{\arraystretch}{1.3}
\caption{The strategies used to evaluate resource allocation.}
\label{tab:strategies}
\centering
\begin{tabular}{ | c | p{2.6in} |}
\hline

\multicolumn{1}{|c|}{Abbr.} &
\multicolumn{1}{|c|}{Resource Allocation Strategy} \\ \hline \hline

ST1 & Always use non-GPU instances \\ \hline
ST2 & Always use GPU instances \\ \hline
ST3 & \textbf{This Paper:} Use non-GPU and GPU instances to reduce the overall 
cost of the instances \\ \hline
 
\end{tabular}
\end{table}

\begin{table}[!t]
\renewcommand{\arraystretch}{1.3}
\begin{center}
\caption{The scenarios used to compare allocation 
strategies.}
\begin{tabular}{ | l | l | r | r |} \hline

Scenario &
Program &
Frame Rate & 
Cameras \\ \hline \hline

\multirow{2}{*}{1}
 & VGG-16 & 0.25 & 1 \\ \cline{2-4}
 & ZF & 0.55 & 3 \\ \hline
\multirow{2}{*}{2}
 & VGG-16 & 0.20 & 1 \\ \cline{2-4}
 & ZF & 0.50 & 1 \\ \hline
\multirow{2}{*}{3}
 & VGG-16 & 0.20 & 2 \\ \cline{2-4}
 & ZF & 8.00 & 10 \\ \hline

\end{tabular}
\label{tab:scenarios}
\end{center}
\vspace{-0.15in}
\end{table}

\begin{table}[!t]
\renewcommand{\arraystretch}{1.3}
\caption{
The types and numbers of instances determined by different allocation 
strategies to handle different scenarios.
}
\label{tab:scenarios_output}
\centering
\begin{tabular}{ | c | r | r | r | r | r |} 
\hline

\multirow{2}{*}{Scen.} &
\multirow{2}{*}{Strategy} &
\multicolumn{2}{|c|}{Instances} &
\multicolumn{1}{|c|}{Hourly} &
\multicolumn{1}{|c|}{Cost} \\ \cline{3-4}

 & & non-GPU & GPU & 
\multicolumn{1}{|c|}{Cost} & \multicolumn{1}{|c|}{Savings}  
\\ \hline \hline

\multirow{3}{*}{1}
 & ST1 & 4 & - & \$1.676 & 0\%\\ \cline{2-6}
 & ST2 & - & 1 & \$0.650 & 61\%\\ \cline{2-6}
 & ST3 & - & 1 & \$0.650 & 61\%\\ \hline\hline
\multirow{3}{*}{2}
 & ST1 & 1 & - & \$0.419 & 36\%\\ \cline{2-6}
 & ST2 & - & 1 & \$0.650 & 0\%\\ \cline{2-6}
 & ST3 & 1 & - & \$0.419 & 36\%\\ \hline\hline
\multirow{3}{*}{3}
 & ST1 & Fail & Fail & Fail & Fail\\ \cline{2-6}
 & ST2 & - & 11 & \$7.150 & 0\%\\ \cline{2-6}
 & ST3 & 1 & 10 & \$6.919 & 3\%\\ \hline
 
\end{tabular}
\vspace{-0.15in}
\end{table}

In order to compare the three strategies, we use the three scenarios described in Table~\ref{tab:scenarios}. The table shows the programs, frame rates, and the number of data streams being analyzed in each scenario. Table~\ref{tab:scenarios_output} shows the types and numbers of instances determined by each strategy to handle each scenario:

\textbf{Scenario 1:}
ST1 uses 4 non-GPU instances to handle the 4 data streams.
That is because a single non-GPU instance can handle only one stream due to the 
high CPU requirement of the programs at these frame rate.
ST2 and ST3 use a single GPU instance to handle all the 4 streams because the CPU 
requirement is decreased significantly while using the GPU.
This saves 61\% of the overall hourly cost compared with ST1.

\textbf{Scenario 2:}
The CPU and GPU requirements of VGG-16 at 0.2 FPS and ZF at 0.5 FPS are 
low such that a single instance can handle the two streams at the same time.
ST1 and ST3 use a single non-GPU instance and either of them saves 36\% of the 
overall hourly cost compared with ST2 which uses a single GPU instance.

\textbf{Scenario 3:}
ST1 fails to execute ZF at 8 FPS since the CPU only can execute ZF at a maximum 
of 0.56 FPS.
ST2 uses 10 GPU instances to handle the 10 data streams of ZF and a single GPU
instance to handle both the 2 streams of VGG-16.
That is because a single GPU instance can handle only one stream of ZF at 8 
FPS due to the high CPU requirement.
ST3 considers both GPU and non-GPU instances to reduce the overall hourly 
cost so it can replace a GPU instance with a non-GPU instance.
Hence, ST3 saves 3\% of the cost compared with ST2.

These experiments demonstrate that different resource allocation strategies are 
best in different scenarios according to several factors, such as analysis 
programs and frame rates.
The strategy used by the resource manager described in this paper considers 
both GPU and non-GPU instances and always has the lowest cost compared with 
the other strategies.

\section{Conclusion}

This paper describes a resource manager that reduces the monetary cost of using 
the cloud to analyze real-time multimedia content from network cameras while 
meeting the desired analysis frame rates.
The manager uses GPU to achieve higher frame rates and considers both GPU and 
non-GPU instances to reduce the overall cost.
The manager formulates the resource allocation problem as a multiple-choice 
vector bin packing problem and solves it using an existing algorithm.
The experiments show that the manager can reduce up to 61\% of the 
cost compared with other allocation strategies.

\section{Acknowledgements}
The authors would like to thank the organizations that provide the camera data.
A complete list of the data sources is available at~\url{https://www.cam2project.net/ack/}.
This project is supported in part by National Science Foundation ACI-1535108, 
CNS-0958487, and OISE-1427808. Any opinions, findings, and conclusions or 
recommendations expressed in this material are those of the authors and do not 
necessarily reflect the views of the sponsors.
\bibliographystyle{IEEEbib}
\bibliography{main}

\begin{thebibliography}{10}

\bibitem{VGG16}
Karen Simonyan and Andrew Zisserman,
\newblock ``Very deep convolutional networks for large-scale image
  recognition,''
\newblock {\em Computing Research Repository}, vol. abs/1409.1556, 2014.

\bibitem{ZF}
Matthew~D. Zeiler and Rob Fergus,
\newblock ``Visualizing and understanding convolutional networks,''
\newblock {\em Computing Research Repository}, vol. abs/1311.2901, 2013.

\bibitem{ChenYangLindnerBarrenetxeaVetterli2012HowisWeatherImageProcessing}
Zichong Chen et~al.,
\newblock ``Howis the weather: Automatic inference from images,''
\newblock in {\em IEEE International Conference on Image Processing}, 2012, pp.
  1853--1856.

\bibitem{srivastava2013video}
Satyam Srivastava and Edward~J. Delp,
\newblock ``Video-based real-time surveillance of vehicles,''
\newblock {\em Journal of Electronic Imaging}, vol. 22, no. 4, pp. 041103,
  2013.

\bibitem{Tian2008}
Ying-li Tian et~al.,
\newblock ``Ibm smart surveillance system (s3): event based video surveillance
  system with an open and extensible framework,''
\newblock {\em Machine Vision and Applications}, vol. 19, no. 5, pp. 315--327,
  2008.

\bibitem{CAM2GlobalSIP}
A.~S. Kaseb et~al.,
\newblock ``A system for large-scale analysis of distributed cameras,''
\newblock in {\em IEEE Global Conference on Signal and Information Processing},
  2014, pp. 340--344.

\bibitem{zhu2011multimedia}
Wenwu Zhu et~al.,
\newblock ``Multimedia cloud computing,''
\newblock {\em IEEE Signal Processing Magazine}, vol. 28, no. 3, pp. 59--69,
  2011.

\bibitem{CAM2CCBD}
A.~S. Kaseb et~al.,
\newblock ``Cloud resource management for image and video analysis of big data
  from network cameras,''
\newblock in {\em International Conference on Cloud Computing and Big Data},
  2015.

\bibitem{4607358}
J.~Fung and S.~Mann,
\newblock ``Using graphics devices in reverse: Gpu-based image processing and
  computer vision,''
\newblock in {\em IEEE International Conference on Multimedia and Expo}, June
  2008, pp. 9--12.

\bibitem{7177522}
C.~Oh et~al.,
\newblock ``Real-time face detection in full hd images exploiting both embedded
  cpu and gpu,''
\newblock in {\em IEEE International Conference on Multimedia and Expo}, June
  2015, pp. 1--6.

\bibitem{4284972}
C.~Y. Lee et~al.,
\newblock ``Multi-pass and frame parallel algorithms of motion estimation in
  h.264/avc for generic gpu,''
\newblock in {\em IEEE International Conference on Multimedia and Expo}, July
  2007, pp. 1603--1606.

\bibitem{6011847}
M.~Alcoverro et~al.,
\newblock ``A real-time body tracking system for smart rooms,''
\newblock in {\em IEEE International Conference on Multimedia and Expo}, July
  2011, pp. 1--6.

\bibitem{Brando201656}
Filipe Brandao and Joao~Pedro Pedroso,
\newblock ``Bin packing and related problems: General arc-flow formulation with
  graph compression,''
\newblock {\em Computers \& Operations Research}, vol. 69, pp. 56 -- 67, 2016.

\bibitem{FasterRCNN}
Shaoqing Ren et~al.,
\newblock ``Faster r-cnn: Towards real-time object detection with region
  proposal networks,''
\newblock in {\em Advances in Neural Information Processing Systems 28},
  C.~Cortes, N.~D. Lawrence, D.~D. Lee, M.~Sugiyama, and R.~Garnett, Eds., pp.
  91--99. Curran Associates, Inc., 2015.

\end{thebibliography}

\end{document}